# Quantum interference in graphene nanoconstrictions

Pascal Gehring[1,*], Hatef Sadeghi[2], Sara Sangtarash[2], Chit Siong Lau[1], Junjie Liu[1], Arzhang Ardavan[3], Jamie H. Warner[1], Colin J. Lambert[2], G. Andrew. D. Briggs[1], Jan A. Mol[1]

[1]Department of Materials, University of Oxford, 16 Parks Road, Oxford OX1 3PH, United Kingdom

[2]Quantum Technology Centre, Physics Department, Lancaster University, Lancaster LA1 4YB, United Kingdom

[3]Clarendon Laboratory, Department of Physics, University of Oxford, Parks Road, Oxford OX1 3PU, United Kingdom

[*]pascal.gehring@materials.ox.ac.uk

**Abstract**

**We report quantum interference effects in the electrical conductance of chemical vapour deposited graphene nanoconstrictions fabricated using feedback controlled electroburning. The observed multi-mode Fabry-Pérot interferences can be attributed to reflections at potential steps inside the channel. Sharp anti-resonance features with a Fano line shape are observed. Theoretical modelling reveals that these Fano resonances are due to localised states inside the constriction, which couple to the delocalised states that also give rise to the Fabry-Pérot interference patterns. This study provides new insight into the interplay between two fundamental forms of quantum interference in graphene nanoconstrictions.**

A key feature of electron transport through single molecules and phase-coherent nanostructures is the appearance of transport resonances associated with quantum interference.[1] Examples include Breit-Wigner resonances, multi-path Fabry-Pérot resonances and Fano resonances. Fano resonances can be observed when a localised state interacts with a continuum of extended states and can lead to very steep gradients in the transmission. Unlike Breit-Wigner resonances, they are not life-time broadened by coupling to the electrodes. The steep slope of Fano resonances makes them attractive for low-power switching and for creating structures with high thermoelectric performance.[2] In what follows, we report the first observation of Fano resonances in electroburnt graphene nanoconstrictions. In addition to these Fano features, the conductance maps exhibit interference patterns which we attribute to multi-mode Fabry-Pérot (FP) interferences. Theoretical modelling reveals that the Fano resonances arise from interaction between the delocalised state giving rise to the Fabry-Pérot pattern and a localised state inside the constriction.

Carbon-based nanostructures, such as metallic or semiconducting single carbon chains[3, 4], graphene nanoribbons and graphene nanoconstrictions are interesting platforms for the study of spintronics[5] and might enable novel technological applications[6]. Graphene nanoconstrictions and nanogaps also provide a robust platform for studying the electric[7], thermoelectric[8] and magnetic[9] properties of single molecules. When they are sufficiently narrow, graphene nanoribbons can be used to build field-effect transistors with an on/off ratio that can exceed 1000.[10] In very narrow constrictions, with a width smaller than the electronic wavelength of electrons, quantum interference effects in analogy to subwavelength optics are predicted[11, 12]. Graphene nanoconstrictions have been fabricated by means of electron beam lithography[13], gold break-junction etching masks[10], local gating[14] and electroburning of graphene[15, 16]. Electroburning has also been used to fabricate graphene quantum dots with addition energies up to 1.6 eV, enabling the observation of Coulomb blockade at room temperature[17]. In this study we use feedback-controlled electroburning to narrow down lithographically-defined bowtie shaped graphene constrictions[18] and study their electronic transport behaviour.

Our devices are fabricated from single-layer CVD-grown graphene[19] which we transfer onto a Si/300nm $SiO_2$ wafer with pre-patterned 10 nm Cr / 70 nm Au contacts. We pattern the graphene into a bowtie shape (see Figure 1a,b) using standard electron beam lithography and $O_2$ plasma etching. The channel length $L$ of the devices and the width $W$ of the narrowest part of the constriction are 4 µm and 200 nm, respectively (see Figure 1a). Our devices are p-doped with a Dirac point $V_{Dirac}$ around 60 V (see Figure 1c). The single-layer nature of the graphene constriction is confirmed by the intensity ratio $I(2D) / I(G)$ » 1 of the Raman G and 2D peaks (see Figure 1d) and the fact that the 2D peak consists of a single

Lorentzian.[20, 21] In addition, we observe a D and D' peak which we attribute to the defective graphene edges formed during the plasma etching.[21] These defect peaks are not present in bulk single-layer graphene samples.[19] To narrow down the constriction we use a feedback-controlled electroburning technique in air, similar to the one described in Ref [18]. We ramp-up a voltage applied between the source and drain contact while monitoring the current with a 5 kHz sampling rate (see Figure 1b). As soon as a drop in the current is detected, the voltage is quickly ramped back to zero. This cycle is repeated until the low bias source-drain resistance of the device, which is measured after each burning cycle, exceeds a threshold resistance of 500 MΩ. The feedback conditions are adjusted for each burning cycle depending on the threshold voltage $V_{th}$ at which the drop in the previous cycle occurred. The current-voltage ($I - V_b$) traces of a typical electroburning process are shown in Figure 1e, where the $I - V_b$ traces before electroburning and after the threshold resistance is reached are coloured blue and red, respectively.

During electroburning, the constriction is narrowed down and as a result the resistance of the device increases. At the final stage, the (only several atoms wide) constriction can break completely and a nanometre sized gap is formed.[12] However, for many devices the threshold resistance is reached before a gap is fully formed. In these cases, narrow graphene constrictions or small graphene islands are left between the mesoscopic graphene leads. Graphene quantum dots formed in this manner have been widely studied[15-17, 22] as a possible platform for room temperature single-electron transistors. In the following we discuss the details of the transport characteristics of empty graphene nanogaps, quantum dots and nanoconstrictions recorded at $T = 4$ K in vacuum (~$10^{-6}$ mbar).

The transport regime which we attribute to an empty gap is characterised by low currents and $I - V_b$ characteristics that can be fitted using the standard Simmons model[23] for tunnelling through a single trapezoidal barrier between source and drain (see Figure 2a). In addition, the $I - V_b$ characteristics show no or a relatively small back gate dependence (see Figure 2b). We find gap sizes of 0.5 – 2.5 nm for these junctions, making them a promising platform for single molecule electronics.[7, 24, 25]

Devices in the weakly coupled quantum dot regime show suppressed current at low bias (see Figure 2c) and characteristic Coulomb diamonds as a function of bias and gate voltage (see Figure 2d). These transport features are indicative of sequential electron tunnelling via a weakly coupled quantum dot between source and drain.[26] From the size of the Coulomb diamonds we can extract addition energies for these quantum dots ranging from 20 to 800 meV, comparable to those found by other groups in similar systems.[13, 15-17] The formation of graphene quantum dots during electroburning process is the result of electron/hole localisation due to charge puddles and/or edge disorder as the graphene channel gets narrower.[27] Theoretical calculations have also shown that localised states can form along the edges of wedge-shaped nanoconstrictions.[28] Furthermore, it is possible that small graphene islands on the order of several nm form, which are only weakly coupled to the graphene leads.[17]

The conductance maps of strongly-coupled devices are dominated by "chess board"-like interference patterns as shown in Figure 2f. In some samples we could observe a transition from this chess board pattern to a Coulomb diamond regime at high positive gate voltages of $\gtrsim$ 40V. This observation is similar to results found in recent studies on short graphene junctions[29, 30] and narrow graphene constrictions[31]. In the latter, the chess board pattern was attributed to interference effects of extended states in the source or drain graphene lead connecting the constriction.[31] In general, interference effects occur on a length scale on the order of the phase coherence length, but can have different origins. If the transport in the graphene sample is diffusive, i.e. when charge carriers are predominantly scattered at random impurities like edge disorder, point defects or charge puddles,[27] the origin of the interference pattern is most likely due to quantum interferences of different random scattering paths (universal conductance fluctuations, UCFs). If the channel length is on the order of or shorter than the mean free path of the carriers (quasi-ballistic transport regime), reflections in the channel result in quasi-periodic multi-mode or collective and periodic single-mode Fabry-Pérot interferences. Carriers can get reflected at the metal contacts[30] or at potential barriers formed by intentional local doping.[32] Whether single- or multi-mode interference is observed strongly depends on the detailed device geometry.[33]

Fabry-Pérot interference effects have previously been observed in 1D nanowires[34], carbon nanotubes[35] and 2D graphene[30], while UCFs have been observed in mesoscopic single-[36], bi- and tri-layer[37] and epitaxial graphene samples[38]. To distinguish between these different types of quantum interference, the chess board conductance patterns need to be carefully analysed for hidden periodicities.[30] From the characteristic energy spacing between single features in the conductance maps and Fast Fourier Transforms (FFTs) of the data shown in Figure 2f (see Figure S6a and b in the Supplementary Information) we can extract a typical energy spacing of 4 – 5 meV. Using a particle-in-a-box approximation[30] we estimate the relevant length scale $L = hv_F/(2E)$ to be between 400 nm for the theoretical local density approximation limit of the Fermi velocity of $v_F = 0.8 \times 10^6$ m/s and 1.1 μm for a Fermi velocity of $v_F = 2.4 \times 10^6$ m/s measured for CVD graphene on a quartz substrate.[39] This length scale corresponds to half the minimal distance over which the electrons remain phase coherent, therefore we can infer a lower bound for the phase coherence length $L_\varphi > 800$ nm in our samples.[29] This value is similar to the value found for exfoliated graphene on $SiO_2$[40], epitaxial graphene[41] and CVD graphene[42]. For short and wide devices small incident angles dominate (longitudinal modes) and resonances appear at $k_F L = n\pi$.[32] However, since our devices are not in the limit $W/L \gg 1$, both longitudinal and transversal modes need to be considered. To model conductance maps for different aspect ratios we have performed nearest-neighbour tight-binding calculations[33] (see section S5 Supplementary Information). Our calculations confirm that for $W \gg L$ a periodic interference pattern with high contrast can be observed. This is due to the fact that the energy of transversal modes $E_W = hv_F/(2W)$ gets negligibly small. The same holds for the 1D limit $W \to 0$, where $E_W$ goes to infinity. In both cases the transport is dominated by longitudinal modes only. In the intermediate multi-mode regime, periodic longitudinal modes can still be observed in the FFT but with much smaller contrast. Since the aspect ratio $W/L$ of our devices is close to unity we expect that the interference pattern shown in Figure 2b will only be quasi-periodic because of multi-mode interferences. Moreover, the fact that the width $W$ of the samples is not constant will cause the transversal modes to become chaotic.[11]

Because the measured chess board pattern is only quasi-periodic, we cannot exclude UCFs as an origin of the observed pattern. UCFs are normally most pronounced at low doping concentrations when the electrochemical potential of graphene is close to the Dirac point.[30] This is unlikely to be the case in our *p*-doped graphene junctions. In addition, the periodicity which we can correlate with the geometry of the device is very similar for all devices investigated in this study, which makes multi-mode Fabry-Pérot interferences a more likely mechanism to explain our data.

Next, we will investigate the microscopic origin of the FP reflections. Based on our assumption for the Fermi velocity (see above) we estimate that carriers are coherently reflected on a length scale of ≲ 1 µm. The visibility/intensity of FP interferences is determined by the reflectance of the potential steps. Unipolar cavities have a small finesse and result in a small visibility ($G_{max} - G_{min}$) / ($G_{max} + G_{min}$) since the conservation of pseudospin suppresses backscattering in graphene.[32] A smooth bipolar potential step like a *pn* junction formed near a metal-graphene contact has a much higher finesse and leads to pronounced resonance pattern.[32] However, since the length scale of less than 1 µm found above is much smaller than the channel length of 4 µm of our devices there need to be additional potential steps inside the graphene channel apart from the metal contacts. From scanning electron microscopy and micro Raman spectroscopy (see Sections S1 and S4 in the Supplementary Information) we can infer that the local hole concentration within a region of several hundreds of nm around the graphene constriction is increased during electroburning. The increase of hole doping of graphene on $SiO_2$ annealed in air was intensively studied and attributed to doping by $O_2$ and moisture and a change in the degree of coupling between graphene and $SiO_2$.[43] This increased p doping can result in the formation of a $pp^+p$ junction in the central region of the devices (see Figure S6c and d in the Supplementary Information). Possible resonance conditions are reflections between the gold contact/the *pn* junction close to the gold contact and the $pp^+p$ junction or reflections within the $pp^+p$ junction which all have a characteristic length scale of several hundreds of nm. This length scale is on the order of the-mean-free path of charge carriers in our devices (see Supplementary Information), which further corroborates our interpretation that the chess board pattern arises from FP interferences rather than scattering at random impurities inside the channel. The visibility of the FP interferences ($G_{max} - G_{min}$) / ($G_{max} + G_{min}$) > 10% is high in our devices, which indicates that the unipolar $p^+p$ interfaces need to have a sharp potential drop with $k_F d \ll 1$, where $d$ is the length over which the carrier density changes.[44] We estimate this length scale by calculating the Fermi vector using $n = k_F^2/\pi$ and the charge carrier concentration $n = C_g^2(V_g - V_{Dirac})^2/e^2$,[30] where $C_g$ is the capacitance of the back gate and $e$ is the elementary charge. For $V_{Dirac}$ = 60 V (see Figure 1c), $d$ is on the order of 3 nm.

We only see interference patterns in nearly fully-burned devices and not directly after the first electroburning steps. We attribute this to the decreasing conductance of the graphene constriction during electroburning, which decreases the denominator in $(G_{max} - G_{min}) / (G_{max} + G_{min})$ and thus increases the visibility of the interferences. Another possible explanation for the onset of interference pattern after electroburning is the recrystallisation of the constriction,[45] which may lead to a higher mean free path that is required for reflections on the µm scale. The interplay between reduced width and reduced carrier density may also increase the factor $\lambda/W$, where $\lambda = hv_F/E$ is the wavelength of the electrons. If this ratio becomes ≳ 3 – 5 the Fabry-Pérot interferences have a high contrast.[11]

We now turn to the sharp anti-resonances in the interference regime as shown in Figure 3a and b (around $V_g$ = -18 V) in some samples (see Supplementary Information for data of other samples). The slope of this anti-resonance feature is different from the slopes of the multi-mode FP interference patterns. Repeated thermal cycling from 4 K to room temperature did not change the slope and position of the feature observed at 4K (see Figure S8). The feature consists of an anti-resonance/resonance double-peak as shown in Figure 3c. This asymmetric curve has a distinct Fano line shape,[46] which is the result of coherent interaction between a localised resonant state with a delocalised background state.[1] Fano resonances have previously been observed in double donor systems in nanoscale silicon transistors[47] and in bundles of single walled CNTs[48]. Fano resonances are also predicted for single molecule systems, where a backbone state is coupled to the leads and a pendant side-group is only coupled to the backbone but not to the leads.[1] In a graphene constriction connected to mesoscopic graphene leads there are delocalised states that give rise to the previously discussed FP pattern, and bound states e.g. localised along the edges due to edge roughness, that give rise to Coulomb blockade at high positive gate voltages close to the Dirac point (see Figure 2f).[31] We attribute the observed Fano resonances to the coherent interaction between these states.

To estimate the coherent coupling strength between the localised and delocalised states in the graphene nanoconstriction, we fit the low bias current – gate voltage ($I – V_g$) traces to the Fano formula:[48, 49]

$$G(\varepsilon) = G_{\text{non}} + G_{\text{res}} \frac{(\varepsilon + q)^2}{\varepsilon^2 + 1}, \tag{1}$$

where $G_{res}$ is the coherent contribution to the conductance, $q$ is the complex Fano factor,[50] $\varepsilon = 2(E - \varepsilon_s)/\Gamma_{Fano}$, $\varepsilon_s$ and $\Gamma_{Fano}$ are the energy and coupling strength of the resonant localised state and $G_{non}$ is the conductance of the non-resonant channel. We model the non-resonant background as the sum of a constant offset $G_{off}$ and a Breit-Wigner peak $A\frac{\Gamma^2}{\Gamma^2+(E-\varepsilon_b)^2}$. This non-resonant background accounts for the conductance peak close to the observed anti-resonance feature. Fits to our data at different bias voltages using Equation (1) are shown as solid lines in Figure 3c. We find for a low bias of $V_b$ = 0.1 mV: $\varepsilon_s$ = -18.3 meV, Re($q$) = 0.3, Im($q$) = 1.1, |$q$| = 1.1 $\Gamma_{Fano}$ = 0.4 meV, and a Breit-Wigner peak at $\varepsilon_b$ = -20.5 meV with a coupling strength of $\Gamma$ = 1.1 meV using a lever arm d$E$/d$V_g$ of 1 meV/V extracted from the slope of the Fabry-Pérot interference pattern as depicted by the dotted black line in Figure 3a. The Fano factor $q$ is a combined measure for the energetic detuning and the ratio of the transmission amplitudes of the resonant and the non-resonant channel.[49] For $q \to \infty$, the transport is dominated by the resonant channel and the line shape becomes that of a Breit-Wigner peak. For $q \to 0$ non-resonant transport dominates resulting in a symmetric dip in the conductance.[49] The value of |$q$| = 1.1 found in our experiments results in an asymmetric feature with characteristic Fano line-shape.[48] The width of $\Gamma_{Fano}$ = 0.4 meV of the resonant state is similar to the values of 0.25 – 0.5 meV found for carbon nanotube bundles.[48] The Fano factor $q$ decreases with increasing positive bias voltage (see inset in Figure 3c) which we attribute to a detuning of the energies of the localised state and the extended states. For large negative bias voltages the detuning changes the Fano factor from 1 to a high value, and the transport is dominated by a resonant channel resulting in a Breit-Wigner peak.

The slope of the Fano feature, as seen in Figure 3a, results from the electrostatic coupling of the localised 'pendant' state to the gate and lead electrodes. Figure 4a, shows a tight-binding model of a pendant state interacting with an extended 'backbone' state. A chain of 5 sites acts as the backbone, while a single site coupled to the second site of the backbone serves as pendant group. Figure 4b shows the calculated transmission coefficient $T(E)$ as a function of energy $E$. A Fano-resonance appears at an energy of about 0.5 eV, which is associated with the site energy of the bound state. The various transmission maxima are Fabry-Pérot resonances of the backbone channel. To calculate the differential conductance characteristic d$I$/d$V_b$($V_b$, $V_g$) of the device for different gate voltages $V_g$, bias and gate voltage dependent transmission coefficients $T(E, V_b, V_g)$ were calculated for two different potential profiles, where i) the bias drops over the left and right contacts (Figure 4c); or ii) the bias drops along the device channel (Figure 4d). In the case where the bias voltage drops across the contacts (see Figure 4c), the on-site energies of the pendant group and the backbone are not influenced by the applied bias voltage. As a consequence the two anti-resonance Fano lines have the same slope as the Fabry-Pérot interference lines (see Figure 4e). In contrast, when the potential drops over the channel (see Figure 4d), the slopes of the anti-resonance lines and the backbone resonances become different (see Figure 4f). As a result of the asymmetry of the junction, one of the Fano lines almost vanishes (see section S7 in the Supplementary Information for details). Comparing the calculations in Figure 4e and f with the experimental data in Figure 3a, we can conclude that, firstly, the investigated junctions are asymmetric and, secondly, that a considerable portion of the applied voltage has to drop across the junction. In a more realistic model, where two hexagonal lattices are connected to various scattering regions with and without pendant groups (see Figure 5), Fano resonances can be only observed in junctions with pendant groups (see section S6 and S8 for more details). Molecular-dynamics simulations and density functional theory calculations of different atomic configurations during nanogap formation[12] further show that dangling carbon atoms and edge disorder can lead to Fano resonance in the transmission spectra of partially burned graphene nanogaps (see section S8 in the Supplementary Information).

In summary we investigated graphene nanoconstrictions fabricated by narrowing down bowtie shaped graphene ribbons using a feedback controlled electroburning technique. In the case of weakly-coupled constrictions, the transport is dominated by Coulomb blockade with addition energies up to 800 meV. In the strongly coupled regime, we observe quasi-periodic chess board like pattern in the conductance maps which we attribute to multi-mode Fabry-Pérot interferences of delocalised states whose length scale agrees with two possible resonance conditions: reflections inside the current-annealed low-doped part of the device or reflections between the electrical contacts and the low-doped part. In some of the devices, we observe sharp anti-resonances features with a Fano line shape inside the interference regime in agreement with our tight binding modelling. We attribute these features to interferences between the extended states and localised states inside the constriction. Such sharp anti-resonances have the potential to underpin the development of low-power switches, because the transmission of the structure can be tuned by a small gate voltage. Moreover, the Mott formula predicts that a high $d\ln G/dV_g$ should also result in a high Seebeck coefficient[51], making such devices promising candidates for thermoelectric energy harvesting.

**Acknowledgements**

We thank the Royal Society for a Newton International Fellowship for J. A. M., the Agency for Science Technology and Research (A*STAR) for a studentship for C.S.L. and a University Research Fellowship for J. H. W. This work is supported by Oxford Martin School, the European Commission (EC) FP7 ITN "MOLESCO" (project no. 606728) and UK EPSRC (grant nos. EP/K001507/1, EP/J014753/1, EP/H035818/1 and EP/J015067/1). This project/publication was made possible through the support of a grant from Templeton World Charity Foundation. The opinions expressed in this publication are those of the author(s) and do not necessarily reflect the views of Templeton World Charity Foundation. The authors would like to thank D. Gunlycke for his help and the useful discussions and Y. Fan and J. Nägele for providing supporting transport data.

**Supporting Information**

Supporting SEM images of the devices, Electrical characterization of graphene used in this study, Supporting micro Raman data, detailed study of interference pattern, additional Fano data, bias dependence of the Fano feature, molecular-dynamics simulations.

**Corresponding Author**

*E-mail: mailto:pascal.gehring@materials.ox.ac.uk

**Captions**

**Figure 1.** (a) False colour SEM image of a graphene constriction (grey) contacted by gold contacts (yellow). (b) Schematic of a graphene nanoconstriction device. (c) Conductance as a function of back gate voltage recorded at $V_b = 100$ mV of an as-prepared device. (d) Raman spectrum of the centre region of the graphene bow-tie after electroburing. (e) $I - V_b$ traces recorded during feedback-controlled electroburning. The first and last traces are shown in blue and red, respectively.

**Figure 2.** Nanostructures with different electronic behaviour formed during electroburning. (a) $I - V_b$ trace and (b) current map of an empty gap. (c) $I - V_b$ trace and (d) current map of a weakly coupled constriction showing sequential tunnelling. (e) $I - V_b$ trace and (f) conductance map of a strongly coupled constriction showing resonance effects. All data was recorded at $T = 4$ K under vacuum. The insets depict a scheme of the constriction.

**Figure 3.** (a) Conductance map at $T = 4$ K of a strongly coupled constriction showing interference effects. A sharp anti-resonance feature around $V_g = -18$V can be observed. The dotted line is used to extract the lever arm. (b) Gate traces for different bias voltages $0.1$ mV $\leq V_b \leq 8$ mV in $0.2$ mV steps of the data shown in (a). The curves are offset by $0.2 \times 10^3 \ e^2/h$ for clarity. (c) Gate traces at different bias voltages (dotted lines) and fits using Equation 1 (solid lines). The inset shows the dependence of the Fano factor $|q|$ as a function of the applied bias voltage.

**Figure 4.** (a) Tight-binding model of a pendant state interacting with an extended 'backbone' state. The backbone is described by a chain of 5 sites with on-site energies $\varepsilon_{1-5}$ that are coupled by hopping matrix elements $-\gamma_{1-4}$ and coupled to the leads via the outer most sites by hopping matrix elements $-\alpha_L$ (on the left side) and $-\beta_R$ (on the right side). The pendant group with an on-site energy $\varepsilon_s$ is coupled to the second site of the backbone by a hopping matrix element $-\alpha$. (b) Calculated transmission coefficient as a function of energy. (c), (d) Sketch of the potential profile where (c) the bias drops over the left and right contacts and (d) the bias voltage drops along the device channel. (e), (f) Corresponding conductance maps as a function of bias and gate voltage for the cases depicted in (c), (d), respectively.

**Figure 5.** Transmission through graphene junctions. (a) Clean graphene ribbon connected to two graphene electrodes, (b-e) graphene junctions with different shape and position of pendent groups. The dotted circles indicate the position of Fano features.

**Figure 1**

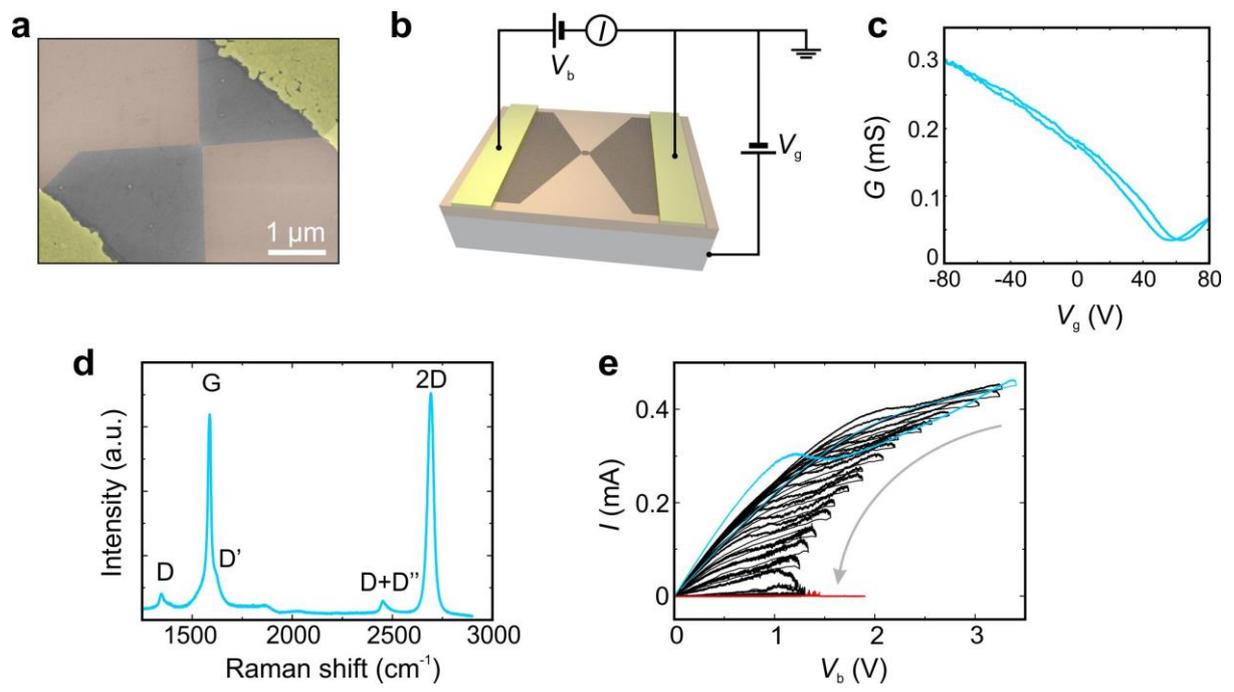

**Figure 2**

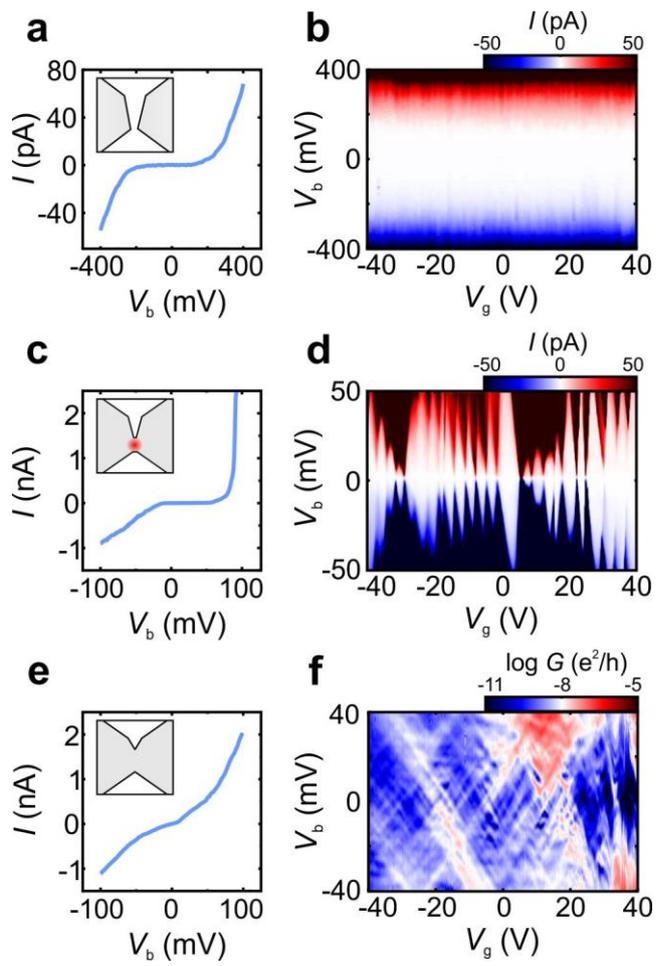

**Figure 3**

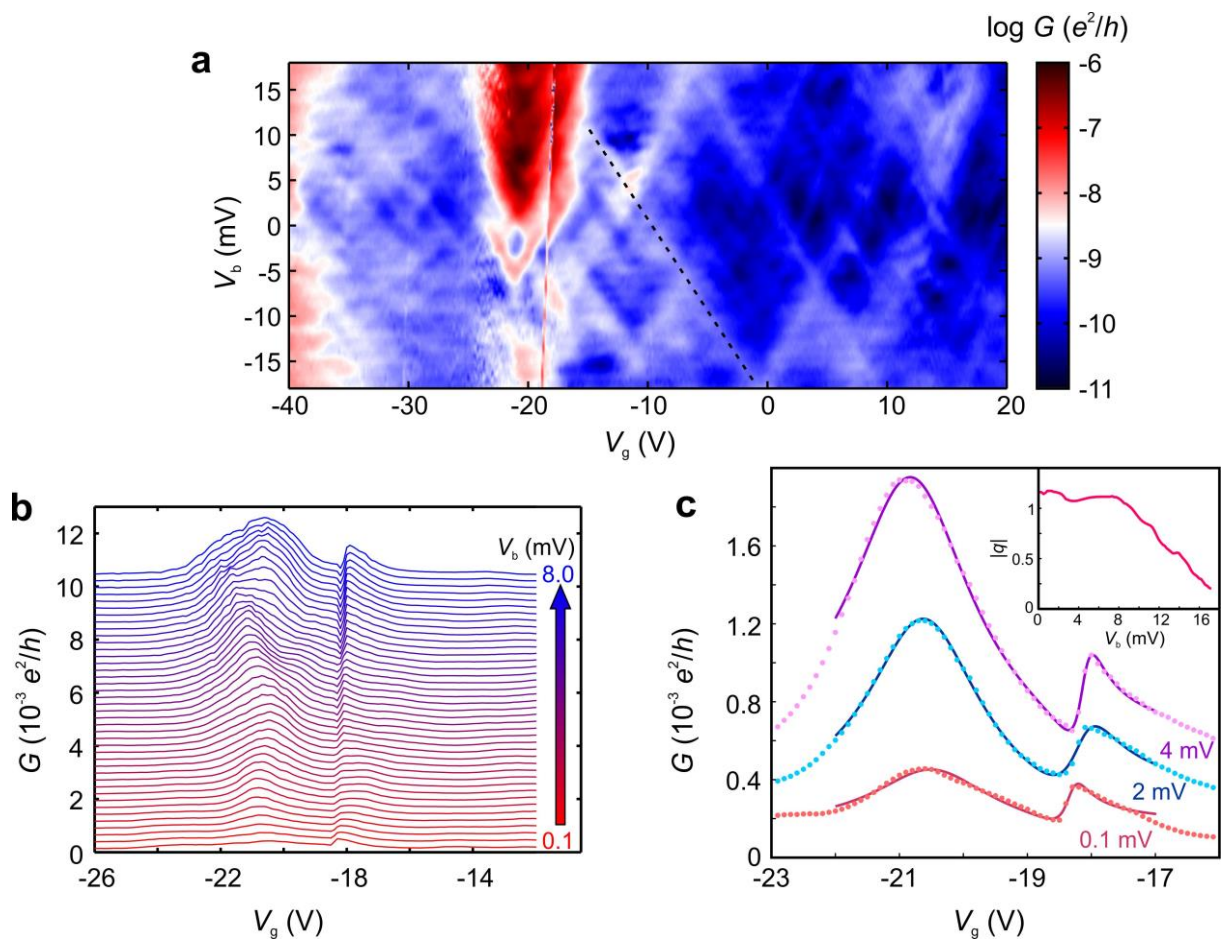

**Figure 4**

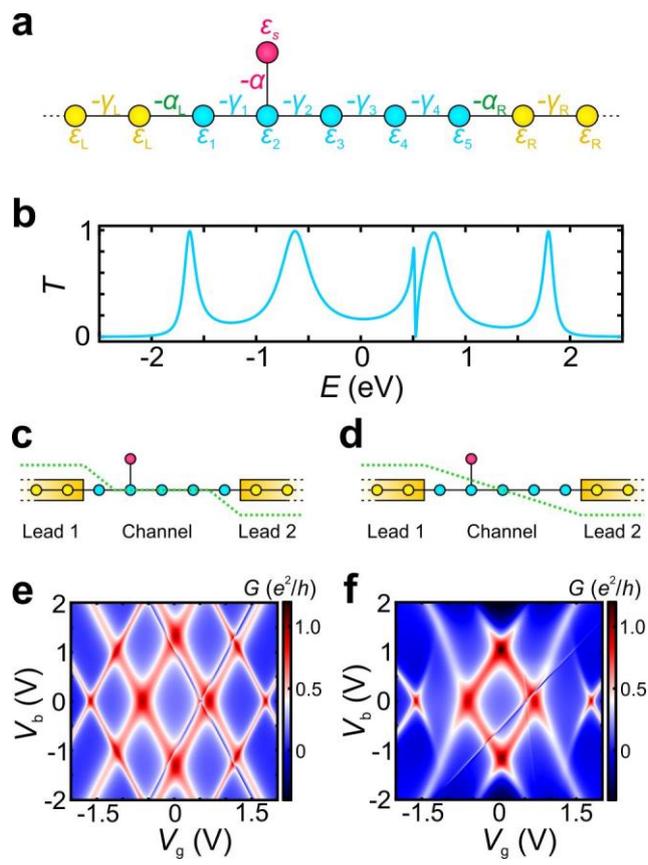

**Figure 5**

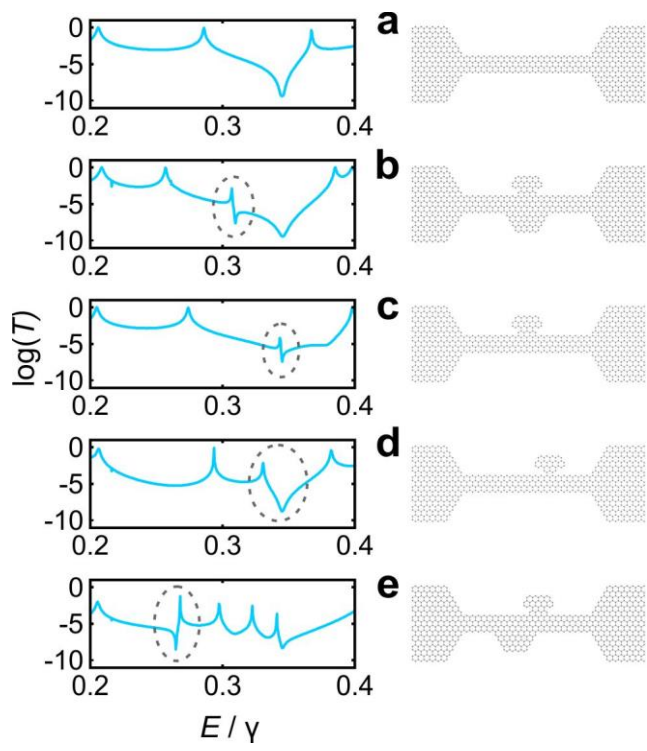

**Supplementary Information**

# Quantum interference in graphene nanoconstrictions

Pascal Gehring[1,*], Hatef Sadeghi[2], Sara Sangtarash[2], Chit S. Lau[1], Junjie Liu[1], Arzhang Ardavan[3], Jamie H. Warner[1], Colin J. Lambert[2], G. Andrew. D. Briggs[1], Jan A. Mol[1]

[1]*Department of Materials, University of Oxford, 16 Parks Road, Oxford OX1 3PH, United Kingdom*

[2]*Quantum Technology Centre, Physics Department, Lancaster University, Lancaster LA1 4YB, United Kingdom*

[3]*Clarendon Laboratory, Department of Physics, University of Oxford, Parks Road, Oxford OX1 3PU, United Kingdom*

[*]pascal.gehring@materials.ox.ac.uk

## S1. Supporting SEM images of electroburned devices

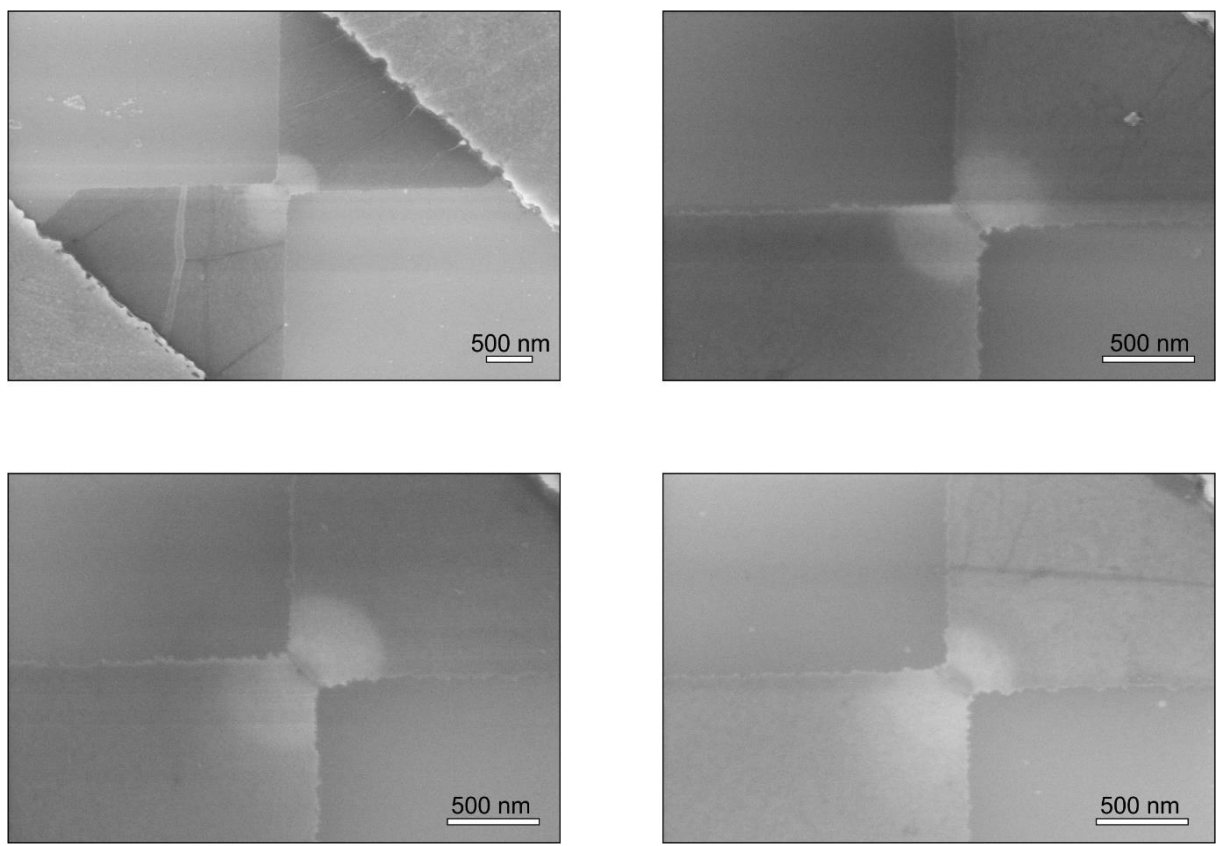

**Figure S1.** SEM images of four different devices after electroburning. In the bright region around the constriction the local hole concentration got increased by current annealing.

Figure S1 shows scanning electron microscopy (SEM) images of four different devices after electroburning. The contrast of the region around the constriction is higher compared to the regions far away from the constriction. There are several contrast mechanisms in SEM like topographic, material/compositional and some special contrast mechanisms like magnetic or electric field contrast. The latter is relevant for *p-n* junctions where local doping creates patch fields above the surface.[1,2] This electric field is positive above the *n-* and negative above the *p*-side of a *p-n* junction, which leads to an acceleration towards (away) the *n* (*p*) side, respectively. As a result, the collection efficiency of secondary electrons above the *p* side is increased, resulting in higher contrast. Thus, the increased intensity close to the graphene constriction observed in Figure S1 can be explained by an increase in local hole doping during electroburning. Additionally, local heating can remove adsorbates which can also lead to a change in contrast.

## S2. Electrical characterisation of the graphene used in this study

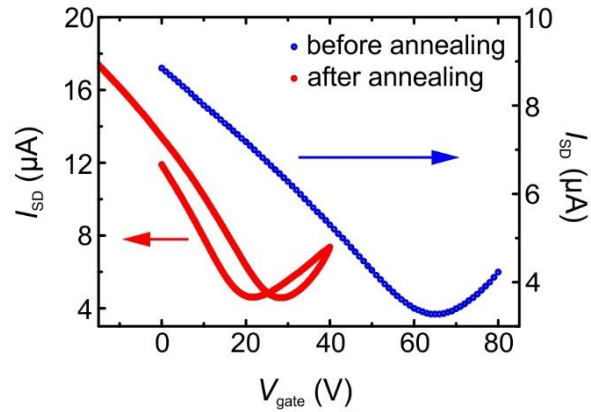

**Figure S2.** Source-drain current as function of back gate voltage ($V_{bias}$ = 50 mV) of an as-prepared device (blue) and a device annealed at 350°C under Ar 75% / $H_2$ 25% for 2h (red).

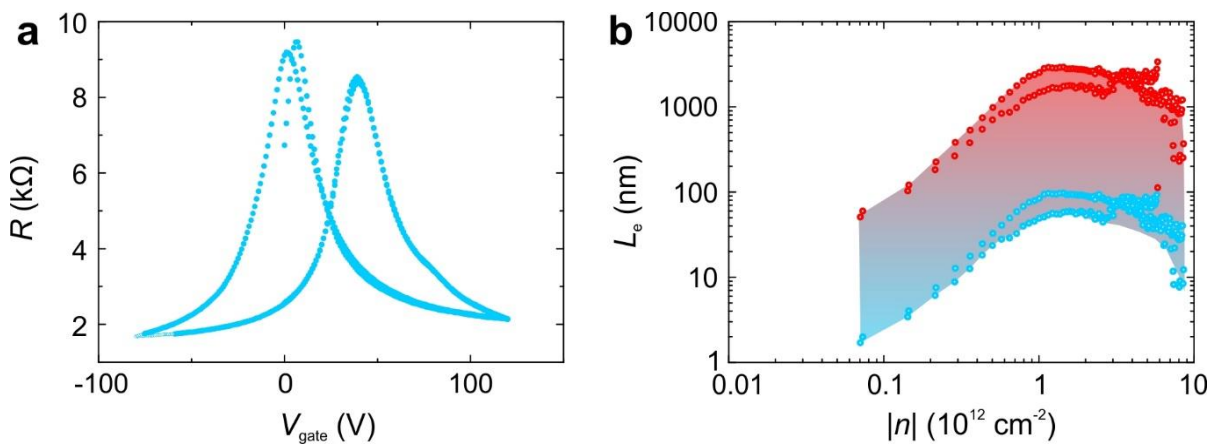

**Figure S3.** (a) Two-terminal resistance as a function of back gate voltage of a partially electroburnt graphene constriction at room temperature under vacuum ($p$ = $10^{-6}$ mbar). (b) Calculated mean-free-path using (a) and assuming a 3000 nm (blue) and 100 nm (red) wide constriction.

The samples used in this study are typically highly *p*-doped. We find that annealing our samples at 350°C under Ar 75% / $H_2$ 25% for 2h efficiently reduces the *p*-doping (see Figure S2). This is most likely due to the removal of residual PMMA used during device fabrication. In addition, keeping the samples under high vacuum for several hours reduces the intrinsic doping (see Figure S3a). Figure S3a shows the two-terminal resistance as a function of back gate voltage of a partially electroburnt graphene constriction at room temperature under vacuum ($p$ = $10^{-6}$ mbar). From this data we can estimate the carrier mobility using the transconductance d$G$/d$V_{gate}$ and:[3]

$$\mu = \frac{dG}{dV_{\text{gate}}} \frac{L}{W C_{\text{ox}}}$$

where the channel length $L = 4$ µm and the back gate capacitance is given by $C_{\text{ox}} = \frac{\varepsilon_0 \varepsilon_r}{d}$ with $\varepsilon_r = 3.9$, $d = 300$ nm. For a channel width of $W = 100$ nm (narrowest part of the constriction defined by lithography) we can estimate $\mu = 38000$ cm²/(Vs). If we neglect the influence of the constriction using $W = 3000$ nm we estimate $\mu = 1300$ cm²/(Vs).

The mean-free-path $L_e$ can be calculated using:[4]

$$L_e = \frac{\hbar}{e} \mu \sqrt{n\pi},$$

where the carrier concentration $n$ is given by the position of the Dirac point $V_{\text{Dirac}}$:

$$n = \frac{C_{ox}(V_{\text{gate}} - V_{\text{Dirac}})}{e}.$$

Figure S3b shows the calculated values of $L_e$ as a function of $n$ using the data in Figure S3a.

## S3. Device characterisation

In total we performed controlled electroburning on 409 graphene junctions. Afterwards we selected 25 devices and recorded their source-drain current $I_{sd}$ as a function of the applied DC bias voltage $V_{sd}$ and back gate voltage $V_g$ (see Figure 1b in the main text) at $T = 4.2$ K in a LHe4 dip-stick setup to distinguish if a device consists of an empty gap, a weakly coupled graphene quantum dot, or a strongly coupled constriction. Of these 25 devices 8 devices consisted of an empty gap, 9 devices were in the weak, and 8 devices were in the strong coupling regime (see discussion in the main text). Neither the absolute room temperature resistance nor the estimated gap sizes obtained by using a Simmons model to fit the *I-V* traces at room temperature could be correlated to the resulting low temperature transport regime.

## S4. Raman study of graphene constrictions

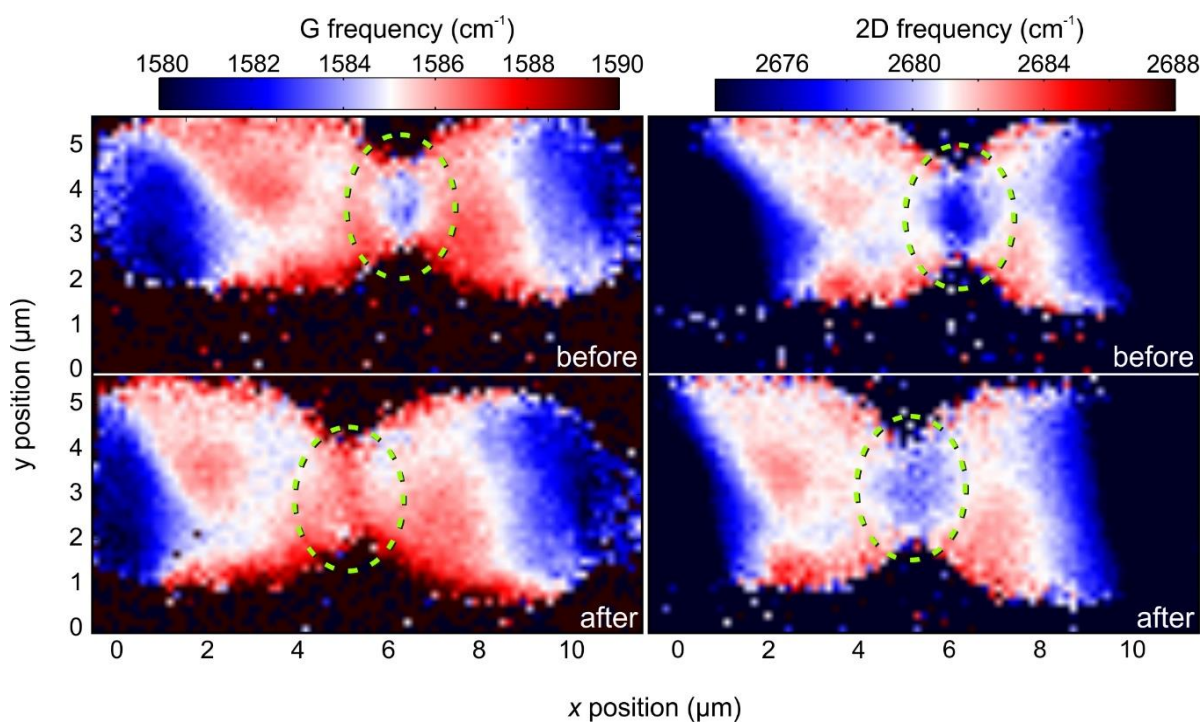

**Figure S4** Raman maps of the G frequency (left) and 2D frequency of untreated (top) and electroburnt (bottom) graphene constrictions. The green dotted lines indicate the position of the constriction.

We performed confocal Raman microscopy measurements of graphene constrictions before and after electroburning. To exclude contact effects we investigated samples with longer channels ($L$ = 10 μm). Measurements were done with a laser wavelength of 533 nm (spot size about 500 nm) and a step size of 150 nm. Figure S4 shows spatial maps of the Raman G (left) and 2D (right) frequency of graphene before (top) and after (bottom) electroburning, obtained by fitting each individual spectrum recorded at each pixel with a Lorentzian function. We observe an increase of the Raman G and 2D frequencies in the constriction area (indicated by the dotted circles) while the rest of the graphene flake stays nearly unaffected. Possible causes for local changes in Raman frequencies are inhomogeneities in the doping level (charge puddles) and differences in strain.[5] To distinguish between strain and doping related effects we followed the correlation analysis suggested by Lee *et al.*[5]. Figure S5a and b show the pixel-to-pixel correlation between the G and 2D frequencies for two different samples. The solid grey and dotted red lines are theoretically calculated[5] equi-strain and equi-doping lines, respectively. The solid red line represents the experimentally determined change in G and 2D frequency as a function of hole doping.[5] The data points fall on a line (see dotted black line) with a slope ($\Delta\omega_{2D}$ / $\Delta\omega_G$) of about 2.2, similar to the value reported in [5]. The highest density of data points is found on or below the zero-strain line, indicating that most of the graphene flake is unstrained or under tensile strain, respectively. In addition, the line describing the data points is parallel to the zero-doping line (dotted grey line), indicating hole doping in our samples. Figure S5c and d show only data points in the constriction are before (purple) and after (green) electroburning. Two changes can be observed: 1) data points move towards the zero-strain line; 2) data points shift along the solid red line towards higher hole doping. These findings can be interpreted as follows:

It was widely observed that before annealing, graphene transferred onto $SiO_2$ substrates is under tensile strain.[5] In our devices, the regions close to the electrical contacts are under high tensile strain most likely due to the fact that our graphene sheets are on top of the 70 nm thick contacts. Although the parts sitting on the substrate are nearly strain free the constriction area is under high tensile strain, most likely due to geometrical effects, i.e. since the constriction is the narrowest part of the channel we expect a local stress enhancement. After electroburning strain in the central region is released indicated by shift of data points towards zero-strain line in Figure S5 c and d. A reduction of tensile strain after annealing at temperatures of >100°C was observed in literature and attributed to annealing-induced slippage and buckling of graphene on $SiO_2$ substrates.[5, 6] In addition, the formation of a crack could result in the local reduction of tensile strain. Interestingly, it can be seen in Figure S5c and d that the data points shift along the zero-strain line towards higher hole doping. The effect of hole doping of graphene on $SiO_2$ annealed in air was intensively studied and attributed to doping by $O_2$ and moisture and a change in the degree of coupling between graphene and $SiO_2$.[6]

In conclusion, electroburning affects the graphene constriction in two ways: it releases the tensile strain and increases *p*-doping locally.

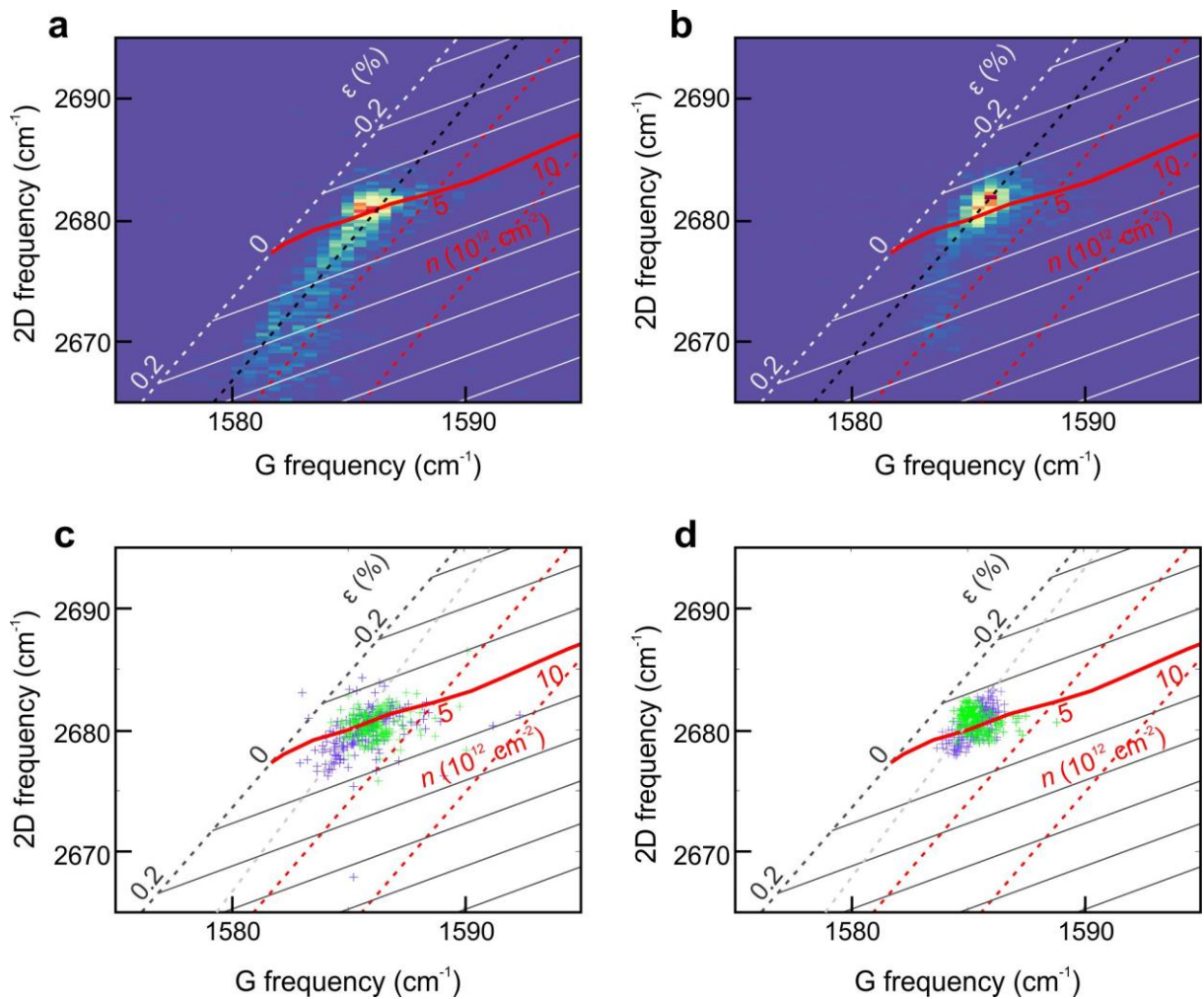

**Figure S5** The effect of hole doping and strain on the Raman G and 2D frequency of graphene. (a), (b) Raman G and 2D correlation for two different devices. The solid grey and dotted red lines indicate lines of equal strain and equal hole doping, respectively. The solid red line is the experimental trajectory for a unstrained graphene sample hole doped using a gate electrode.[5] (c), (d) Data points from the constriction area before (purple) and after (green) electroburning.

## S5. Detailed study of the interference pattern

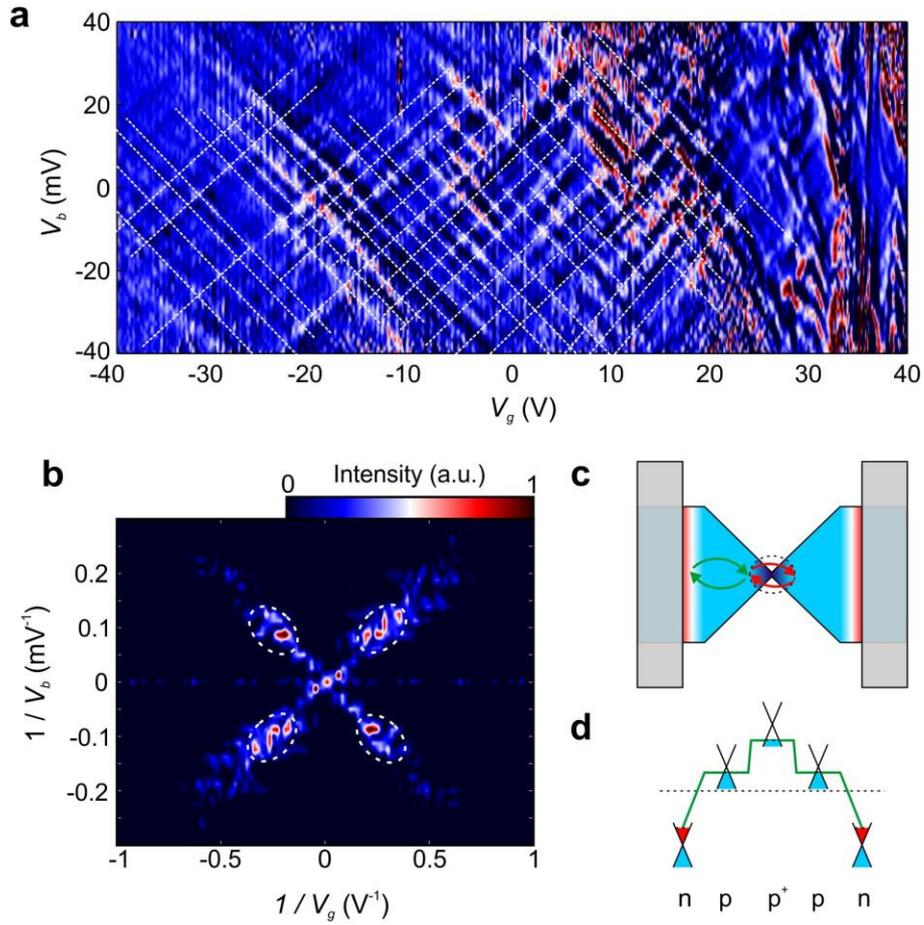

**Figure S6.** (a) Derivative of the conductance map shown in Figure 2f (main text) with respect to $V_g$. The white dashed lines indicate the positions of resonance features. (b) Fourier transform of (a). The hot spots inside the white dotted circles are used to estimate the characteristic energy spacing. (c) Schematic of possible resonance conditions: carriers can be reflected between the *pn* junction formed at the contacts and the highly doped central region (green arrows) or inside the highly doped central region (red arrows). (d) Schematic of the charge carrier distribution inside the channel where the dotted line depicts the charge neutrality point.

In this section we try to find the origin of the quasi periodic interference pattern found in partially burned graphene samples. Those pattern can be attributed to interferences between multiple scattering paths (universal conductance fluctuations (UCF)) or between parallel transport channels reflected at interfaces like *pn* junctions formed at the electrical contacts (multi-/single-mode Fabry Pérot interferences). To differentiate between UCFs and multi-mode Fabry Pérot interferences is not easy since both can result in chaotic/quasiperiodic interference pattern. To get further inside we try to find hidden periodicities in our data.

In Figure S6a we show the derivative of the conductance map presented in Figure 2f (main text) with respect to the back gate voltage $V_g$. This method enhances the contrast of the interference pattern.[7] We added guidelines for the eye (dotted white lines in Figure S6a) that help to identify the position of the interference features. Interestingly, the spacing between those lines does not vary drastically and is on the order of 4 meV. A better way to quantitatively analyse the data is to calculate the Fast Fourier transform (FFT) of Figure S6a.[7] The result is depicted in Figure S6b. We observe regions of high intensity at an energy of about 4 – 5 meV (see dotted circles in Figure S6b and using the bias values of the regions of high intensity). Interestingly we find similar energies for other devices which we attribute to the universal geometry of our samples. This energy corresponds to a length scale of about $L = h v_F/(2E) = 1.1$ μm using $v_F = 2.4 \cdot 10^6$ m/s,[7] which is much smaller than the total length of 4 μm of the devices. Since the width of the bow-ties is not constant it is unlikely that transversal modes can form a regular interference pattern. We find that during electroburning a small region of about 500 nm x 500 nm to 1 μm x 1 μm (we estimated this size by spin coating a thin layer of PMMA on a graphene device and measuring the size of the hole formed in the PMMA after current annealing with an atomic force microscope and by recording SEM images of the devices after electroburning as depicted in Figure S1) around the constriction is current annealed (dotted circle in Figure S6c). This efficiently increases the local hole doping (see Section S4) resulting in a lateral $pp^+p$ junction (see Figure S6d). These changes in doping concentration result in potential steps that can back scatter carriers.[8] Thus, one possible resonance condition is the reflection from the $pp^+$ interfaces as depicted by the red arrows in Figure S6c. It was also reported that carriers can be reflected from the $pn$ junction formed close to the metal contacts[8] (see Figure S6d). Consequently, a second possible resonance condition is the reflection between the contacts and the $pp^+$ interface as depicted by the green arrows in Figure S6c. Using the total channel length of 4 μm and assuming a depleted central region of a length of about 1 μm the condition depicted by the green arrows in Figure S6c corresponds to a length scale of about 1.5 μm, whereas the reflection in the highly doped area should have a characteristic length scale of about 1 μm. Thus, there are at least two possible resonant conditions that might lead to Fabry-Pérot interferences.

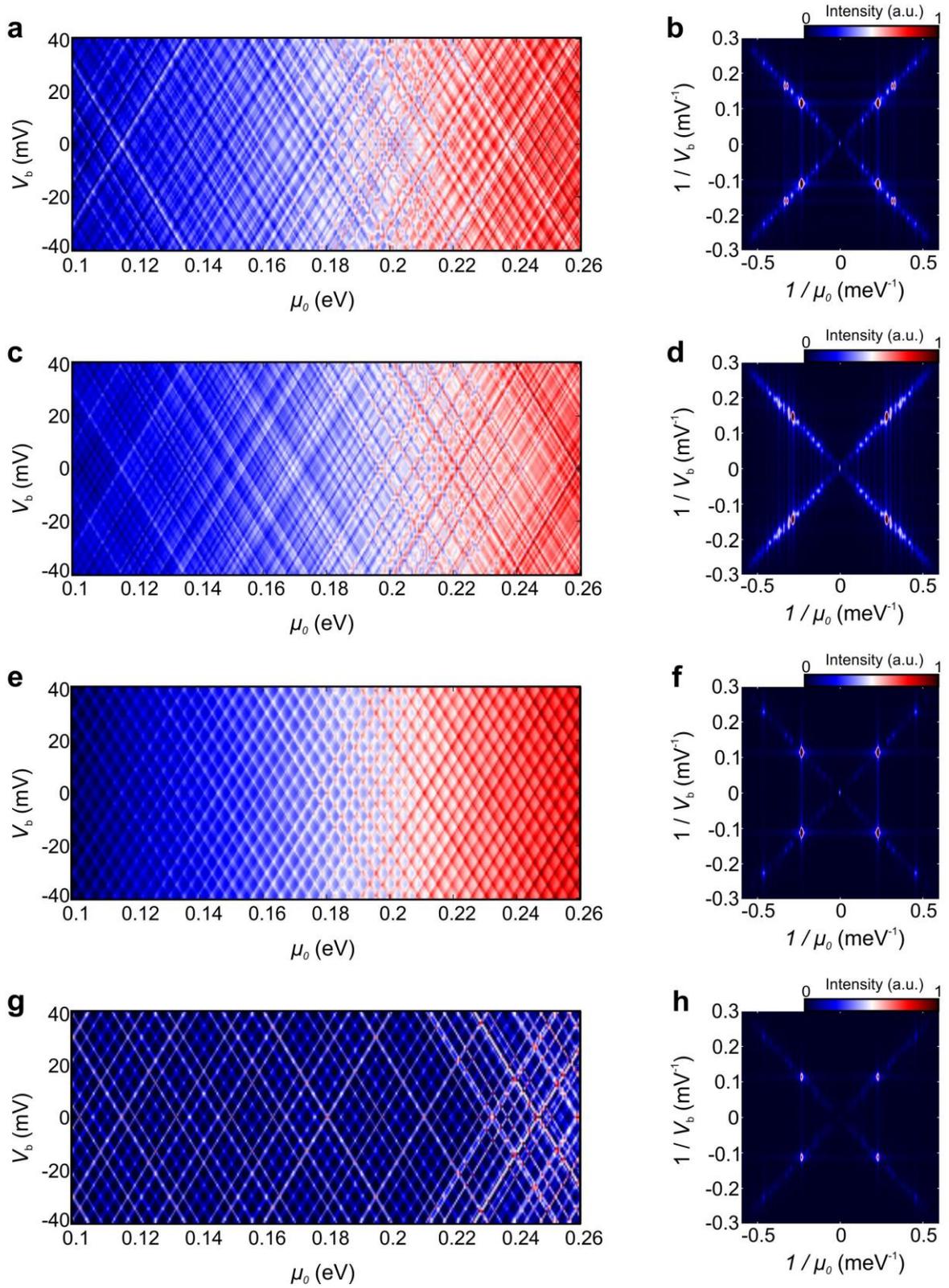

**Figure S7.** Results of the nearest-neighbour tight-binding simulation. (a),(b) Conductance map and Fourier transform of the derivative of the conductance map with respect to the energy $\mu_0$ for (a),(b): $hv_F/(2L) = 4.4$ meV and $hv_F/(2W) = 4.4$ meV, (c),(d): $hv_F/(2L) = 0.4$ meV, $hv_F/(2W) = 4.4$ meV, (e),(f): $hv_F/(2L) = 4.4$ meV, $hv_F/(2W) = 0.4$ meV, (g),(h): $hv_F/(2L) = 4.4$ meV, $hv_F/(2W) = 225$ meV.

In the following we compare our experimental data to interference pattern calculated using a nearest-neighbour tight-binding calculation.[9] We assume highly reflective contact interfaces by choosing a small density of states parameter $\rho_0 = -0.1/(\pi\gamma)$. The resulting visibility of the resonance feature is on the order of 20 – 30 %, very similar to the visibility found in our experiments. As an input parameter we use the same energy scale found in our experiments. The conductance maps where obtained from the calculated transmission function $T(E)$ by $\frac{dI}{dV_b} = \frac{e^2}{h}\left[T\left(\frac{eV_b}{2} - \mu_0\right) + T\left(-\frac{eV_b}{2} - \mu_0\right)\right]$, where $\mu_0$ is the electrochemical potential of the device. The results for three different device geometries with $W/L = 1$, $W/L < 1$ and $W/L > 1$ are depicted in Figure S7a, c and e, respectively. In these simulations we chose either the energy of the longitudinal modes $E_L = hv_F/(2L)$ or the energy of the transversal modes $E_W = hv_F/(2W)$ to be equal to 4.4 meV. We applied the same Gaussian smoothing we used for the calculation of $dI/dV$ of our measured data to the simulated conductance maps to account for high frequency filtering in the Fourier transforms. The simulated conductance maps exhibit interference pattern similar to those found in our experimental data. To quantify their periodicity we performed Fourier transformations of the derivative of the simulated conductance maps with respect to the electro chemical potential $\mu_0$ as depicted in Figures S7 b,d and f. The FFTs show hot spots at similar energies like found in our experiments. These hot spots are more pronounced for aspect ratios $W/L \gg 1$ (see Figure S7e and f) which is the single mode limit where high frequencies due to transversal modes become smaller than $kT$. However, the hot spots are still visible when keeping $L$ constant and decreasing $W$ (see Figure S7a) like it is the case in our bow-tie shaped graphene junctions. For long and narrow ribbons with $W/L < 1$ the interference pattern gets chaotic (multi-mode regime): In Figure S7c and d we assume that the typical energies found in our experiments are due to transversal modes by keeping $E_W = 4.4$ meV. The hot spots in the FFT are still visible but have lower contrast. Only in the 1D-limit where the width of the channel gets small enough so that only a few transversal modes with high energy can form the longitudinal modes become clearly visible again (see Figure S7g and h). We can conclude that interferences due to longitudinal modes as well as transversal modes originating from carrier reflections on a typical length scale of about 1.1 μm can explain our data. Regarding our device geometry (bow-tie with varying $W$) we expect that longitudinal modes are responsible for the periodic resonances. The varying width result in chaotic multi-mode interference pattern which can be seen as lines in the FFTs. It is worth to mention that the Gaussian smooth might allow us to get rid of some high frequency multimode features and thus enhances the contrast of longitudinal modes.

## S6. Supporting Fano resonance data

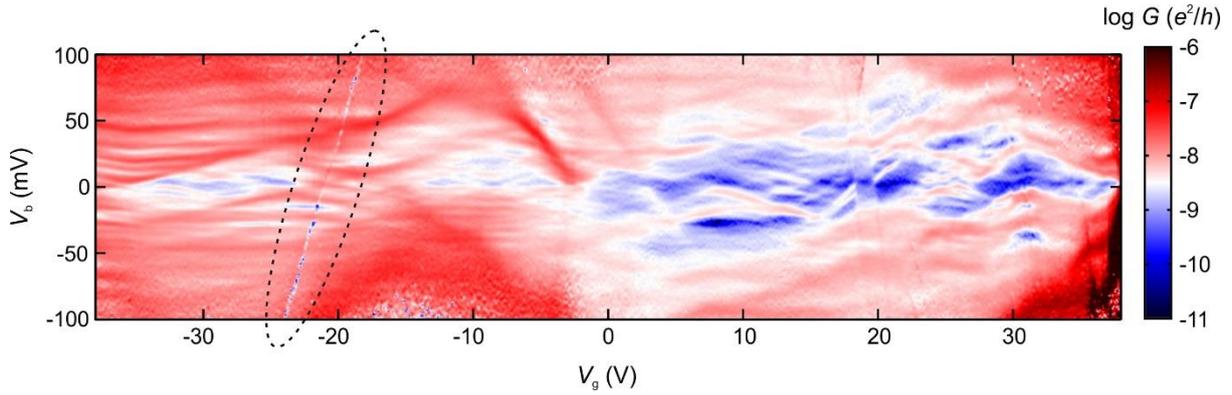

**Figure S8.** Data of the sample discussed in the main text before thermal cycling. During thermal cycling the sample was warmed up to room temperature, exposed to ambient pressure, evacuated and cooled down to 4 K. The sharp resonance feature is highlighted by a dotted line.

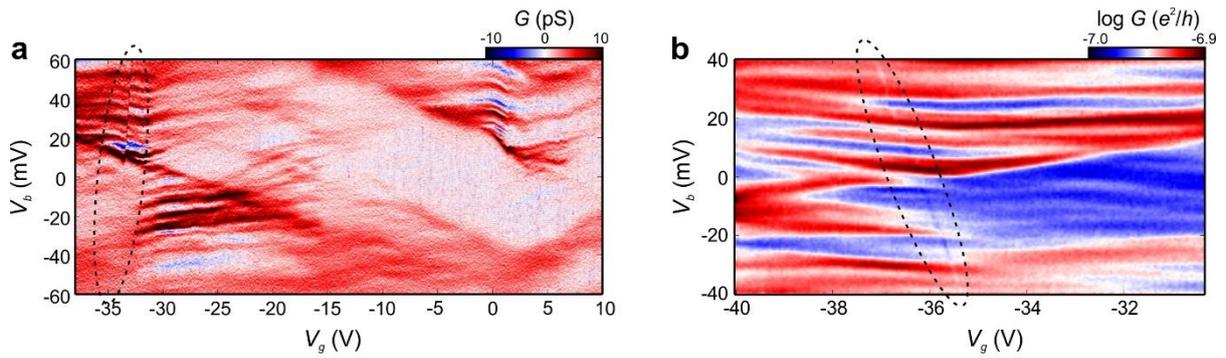

**Figure S9.** Additional data of samples showing sharp resonance features (indicated by the dotted lines).

Figure 5 in the main text shows the transmission coefficient for a series of junctions where a graphene nano-ribbon with or without pendent group is connected covalently to the graphene electrodes. To calculate the transmission coefficient $T(E)$ for electrons with energy $E$ traversing from one electrode to another through these junctions, we use our NEGF method implemented in Gollum.[10] The tight-binding Hamiltonian of each structure is constructed by considering only the $\pi$-orbitals where each atom modelled by one orbital per site with on-site energy zero is coupled to the nearest neighbours with energy integral $\gamma$. When a perfectly clean nanoribbon is connected to the graphene electrodes (Figure 5a), the transmission spectra shows resonances and anti-resonances. To demonstrate that the presence of a pendant group which is weakly coupled to the continuum states of graphene nanoribbon could potentially create Fano-resonance in the transmission spectra, we examined also several junctions where a graphene constriction is attached to the central (perfect) ribbon. As shown by black dashed circles in Figure 5, Fano-resonances are created in the presence of a pendent group (Figure 5b-e).

## S7. Bias dependence of the Fano feature

To calculate the differential conductance characteristic d$I$/d$V_b$$(V_b, V_g)$ of the device for different gate voltages $V_g$, bias and gate voltage dependent transmission coefficients $T(E, V_b, V_g)$ were calculated for two different potential profiles, where i) the bias drops over the left and right contacts (Figure 4c main text); or ii) the bias drops along the device channel (Figure 4d main text). In the first case, neither the bound state nor the most of the channel feel any changes in the bias voltage. Consequently, the bias voltage simply shifts the energy levels of the right and left leads by $-V_b/2$ and $V_b/2$, respectively. In the second case, both the on-site energy of the bound state and backbone are altered by the bias voltage. The energy levels of the left and right lead still change by $-V_b/2$ and $V_b/2$, respectively, and in addition, a linear potential profile $U_b(n) = V_b(3-n)/4$ with $n = [1, 2, 3, 4, 5]$ is added to the on-site energy $\varepsilon_n$ of the site $n$ to model the voltage drop through the junction, where $n = 2$ for the pendant group. For both cases, a constant gate voltage $V_g$ is applied only to the channel. The $V_b$ dependent current $I(V_b, V_g)$ is calculated using the Landauer formula:

$$I(V_b, V_g) = \frac{2e}{h} \int_{-\infty}^{+\infty} dE \, T(E, V_b, V_g) \left( f\left(E + \frac{eV_b}{2}\right) - f\left(E - \frac{eV_b}{2}\right) \right). \tag{2}$$

$G_{\text{diff}}$ is calculated by evaluating the integral in equation (2), and subsequently differentiating the result with respect to the bias voltage:

$$G_{\text{diff}}(V_b, V_g) = \frac{dI(V_b, V_g)}{dV_b}. \tag{3}$$

This method can lead to long calculation times, which can be reduced by considering only the first term of the Taylor expansion of the Fermi-Dirac distribution function $f(E)$ in equation (2). Therefore, a simplified and approximated form of $G_{\text{diff}}$ is obtained in the limit of low temperatures:

$$\hat{G}_{\text{diff}} = G_0 \frac{T(E_F = V_b/2, V_b, V_g) + T(E_F = -V_b/2, V_b, V_g)}{2}, \tag{4}$$

where $G_0 = e^2/h$ is the conductance quantum. We calculate both conductances, $G_{\text{diff}}(V_b, V_g)$ and $\hat{G}_{\text{diff}}(V_b, V_g)$, by calculating the transmission coefficients $T(E, V_b, V_g)$ and using equation (3) and (4), respectively.

Figure 4e and f show conductance maps $G_{\text{diff}}(V_b, V_g)$ calculated using the exact method (equation (3)).

The slope of the Fano line shape determines the amount of the bias voltage felt by the bound state in the graphene constriction. In addition, the relative change in contrast of the Fano lines in the conductance maps determines how asymmetric the junction is. This performs like an indirect telescope to image the physical properties of the junction.

It is worth to mention that both models, equation (3) and (4), are able to describe the effect of the voltage drop on the slope of the Fano feature. However, the influence on the intensity of the feature can be captured only if the "exact method" (equation (3)) is used. (for a comparison see Figure S10 and S11).

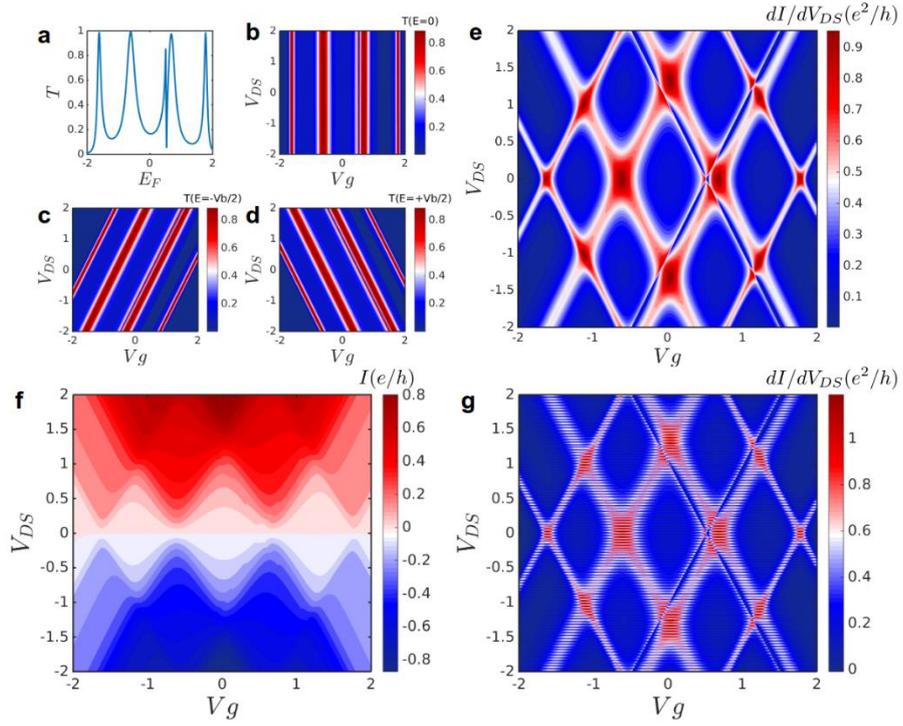

**Figure S10.** Differential conductance calculation for the structure shown in Figure 4c where most of the voltage drops across the contacts (a-e) using equation (3) in the main text ("approximation method"), and (f-g) using equation (4) in the main text ("exact method"). (a) Transmission coefficient $T(E)$, (b) $T(E_F, V_b, V_g)$ at $E_F=0$, (c) $T(E_F, V_b, V_g)$ at $E_F=-V_b/2$, (d) $T(E_F, V_b, V_g)$ at $E_F=+V_b/2$, (e) $\hat{G}_{\text{diff}}(V_b, V_g)$ calculated using equation (4) ("approximation method"), (f) $I(V_b, V_g)$ calculated using equation (3) ("exact method") and (g) $G_{\text{diff}}(V_b, V_g)$ using the "exact method".

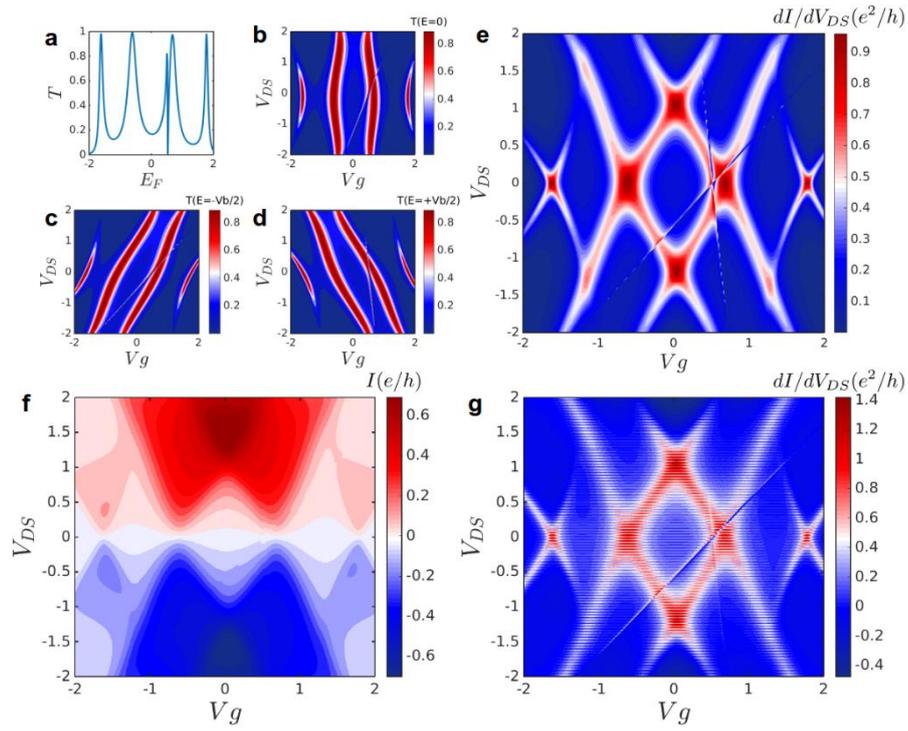

**Figure S11.** Differential conductance calculation for the structure in Figure 4d, where the voltage drops mainly across the junction (a-e) using equation (3) in the main text ("approximation method"), and (f-g) using equation (4) in the main text ("exact method"). (a) Transmission coefficient $T(E)$, (b) $T(E_F, V_b, V_g)$ at $E_F=0$, (c) $T(E_F, V_b, V_g)$ at $E_F=-V_b/2$, (d) $T(E_F, V_b, V_g)$ at $E_F=+V_b/2$, (e) $\hat{G}_{\text{diff}}(V_b, V_g)$ calculated using equation (3) ("approximation method"), (f) $I(V_b, V_g)$ calculated using equation (2) ("exact method") and (g) $G_{\text{diff}}(V_b, V_g)$ using "exact method".

**S8. Fano features in randomly formed graphene junctions**

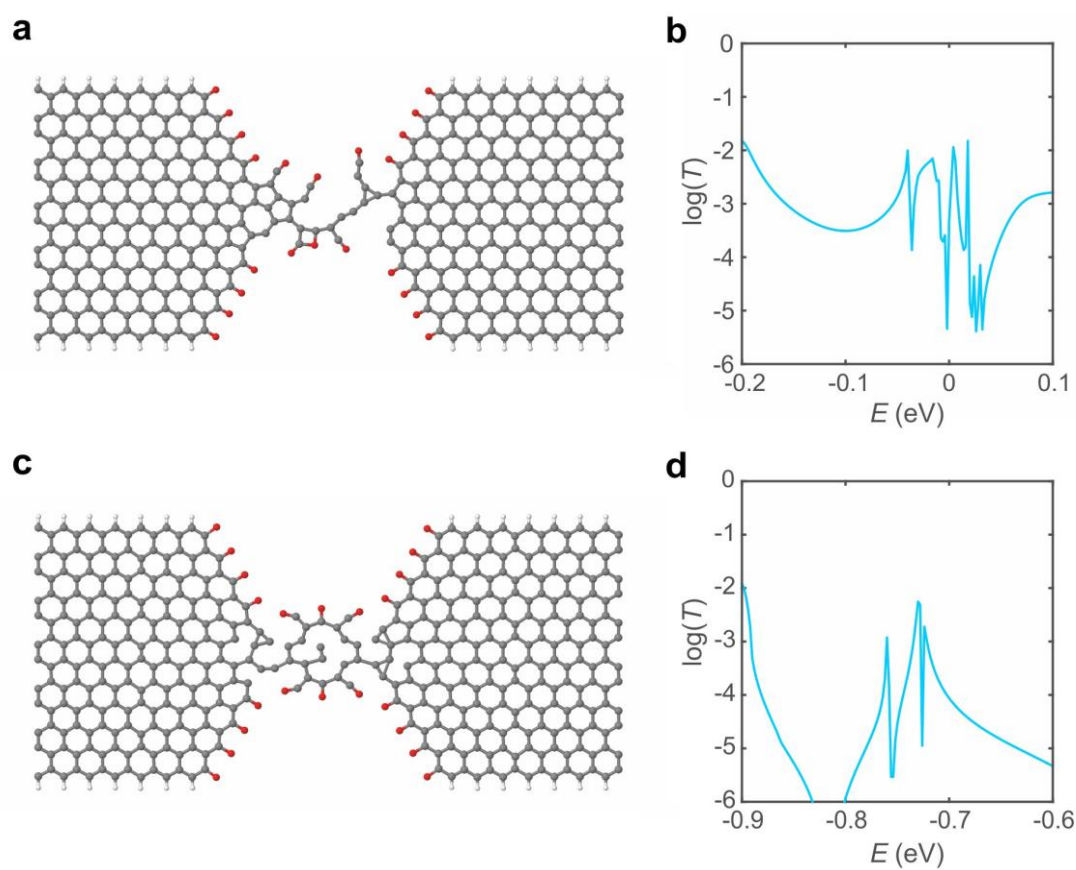

**Figure S12.** Atomic configuration of a junction with (a) one and (c) two pathways and with oxygen (red), hydrogen (white) as well as carbon (grey) terminations modelled by classical molecular-dynamics simulations. (b) and (d): Transmission probability of the structure shown in (a) and (c), respectively, calculated using DFT combined with NEGF methods.[11]

**Supporting references:**